\documentclass[aps,prd,preprint,superscriptaddress,showpacs,floatfix,nobibnotes,nofootinbib]{revtex4-1}
\usepackage[usenames, dvipsnames]{color}

\usepackage{float}
\usepackage{latexsym}
\usepackage{amsmath}
\usepackage{amssymb}
\usepackage{graphicx}
\usepackage{hyperref}
\usepackage{setspace}
\usepackage{longtable}
\usepackage{enumitem}
\usepackage[bottom]{footmisc}

\usepackage[normalem]{ulem}
\usepackage[usenames, dvipsnames]{color}

\usepackage{setspace}
\usepackage{comment}

\usepackage{bm}
\usepackage{epsfig}
\newcommand{\bea}{\begin{eqnarray}}
\newcommand{\eea}{\end{eqnarray}}

\newcommand{\nn}{\nonumber}

\begin{document}
\setlength{\baselineskip}{20pt}

\title{Isotropization of a rotating and longitudinally expanding $\phi^4$ scalar system}

\author{Margaret E. Carrington}
\affiliation{Department of Physics, Brandon University,
Brandon, Manitoba R7A 6A9, Canada}
\affiliation{Winnipeg Institute for Theoretical Physics, Winnipeg, Manitoba, Canada}

\author{Gabor Kunstatter}
\affiliation{Department of Physics, University of Winnipeg, Winnipeg, Manitoba, R3M 2E9 Canada}
\affiliation{Department of Physics, Simon Fraser University, Burnaby, British Columbia, V5A 1S6 Canada}
\affiliation{Winnipeg Institute for Theoretical Physics, Winnipeg, Manitoba, Canada}

\author{Christopher D. Phillips}
\affiliation{Department of Physics, Brandon University,
Brandon, Manitoba R7A 6A9, Canada}
\affiliation{current address: Department of Electrical and Computer Engineering, University of Waterloo, Ontario, Canada}

\author{Marcelo E. Rubio}
\affiliation{Department of Physics, Brandon University,
Brandon, Manitoba R7A 6A9, Canada}
\affiliation{Winnipeg Institute for Theoretical Physics, Winnipeg, Manitoba, Canada}
\affiliation{SISSA, 34136 Trieste, Italy and INFN (Sezione di Trieste)}

\date{October 08, 2022}

\begin{abstract}

We present numerical simulations for the evolution of an expanding system of massless scalar fields with quartic coupling.
By setting a rotating, non-isotropic initial configuration, we compute the energy density, the transverse and longitudinal
pressures and the angular momentum of the system. We compare the time scales associated with the isotropization
and the decay of the initial angular momentum due to the expansion, and show that even for fairly large initial angular
momentum, it decays significantly faster than the pressure anistropy.

\end{abstract}

\maketitle


\section{Introduction}

In this paper we study the time evolution of an expanding system of rotating massless real scalar fields with quartic coupling. Our calculation is based on the method developed in \cite{Dusling10,Dusling:2012ig}.
Observables calculated in a loop expansion exhibit divergences at next-to-leading order, which originate from instabilities in the classical solutions. The effect is seen in a calculation of the energy-momentum tensor at next-to-leading order, where the energy density and pressures of the system diverge rapidly with increasing time. 
Gelis {\it et al.} have shown that this problem can be cured using a resummation scheme that collects the leading secular terms at each order of an expansion in the coupling constant. 
This resummation can be done by allowing the initial condition for the classical field to fluctuate, and averaging over these fluctuations. They have shown that a system of scalar fields isotropizes when this resummation is performed \cite{Dusling:2012ig}. 

The motivation behind the development of this approach is to study the thermalization of the glasma phase of the matter created in a relativistic heavy ion collision. It is known that a hydrodynamic description, which is valid when the system is fairly close to thermal equilibrium, 
works well at very early times ($\sim 1$fm/c).
Approaches that are based on kinetic theory descriptions of the scattering of quasi-particles cannot explain this rapid thermalization. Another possibility that has been studied extensively is that the system is strongly coupled, even at very high energies. The proposal of Gelis {\it et al.} is that rapid thermalization could be achieved by a resummation of quantum fluctuations.  
The Colour Glass Condensate (CGC) effective theory provides a natural framework for this formulation \cite{McLerran_1994_a,McLerran_1994_b,McLerran_1994_c}. 
At very early times the system is best described as a system of strong classical fields, that can be obtained from solutions of the Yang-Mills equation using a CGC approach.  
The spectrum of quantum fluctuations was derived in \cite{Epelbaum13PRD}. 
The success of the resummation method was demonstrated in \cite{Epelbaum13PRL}, where the authors showed that pressure isotropiztion occurs in an SU(2) analogue of QCD. 

Our ultimate goal is to use the Gelis {\it et al.} approach to study the creation and evolution of angular momentum in a glasma. This is interesting in the context of recent proposals that the glasma is produced in a rapidly rotating state, which could be detected by looking for the polarization of produced hyperons. There have been calculations that predict very large values for the initial angular momentum of the system  \cite{Gao:2007bc, Becattini:2007sr,Liang:2019clf}, but significant final state polarization effects have not been observed \cite{Adam_2018,Acharya_2020}. In this paper we develop a formulation to calculate the angular momentum of a system of real scalar fields. We present preliminary results that indicate the angular momentum relaxes to a small value on a time scale significantly smaller than the time scale for  pressure isotropization. 
If a similar result is obtained in a QCD glasma, it would be consistent with the observations in \cite{Adam_2018,Acharya_2020}.
We also comment that a calculation of angular momentum in glasma was done in \cite{Carrington_2022_a}, using a CGC approach with a proper time expansion, and found also that large amounts of angular momentum was not produced.

Since computations in a gauge theory are considerably more complicated, we will work with a scalar theory. 
While it is true that QCD and scalar $\phi^4$ theory are different in many ways, they have important similarities in the context of this calculation because they both have unstable modes and are scale invariant at the classical level. In addition, we will 
minic the kinematics of a relativistic nuclear collision by 
working in Milne coordinates with a rapidity independent background field.
Milne coordinates are suitable because in a nuclear collision, there is a preferred spatial direction provided by the collision axis, and in the high energy limit one expects invariance under Lorentz boosts in the $z$-direction. 


This paper is organized as follows. 
In section \ref{sec-formalism} we describe the method, and in section \ref{sec-observables} we formulate the calculation of the energy-momentum tensor and angular momentum.
Some details of our numerical procecure are discussed in section \ref{sec-numerics}. 
In section \ref{sec-results} we present our results, and in section 
\ref{sec-conclusions} we make some concluding remarks. 

Throughout this paper, the spacetime is always taken to be Minkowski, with the signature $(+,-,-,-)$. 
In addition to standard inertial coordinates $(t,x,y,z)$, we will also use Milne coordinates $(\tau,x,y,\eta)$, where $\tau$ is proper time and $\eta$ is spacetime rapidity. Finally, we choose units such that $c=k_B=\hbar=1$, where $c$ is the speed of light in vacuum, $k_B$ is the Boltzmann constant, and $\hbar$ is the Planck constant divided by $2\pi$.

\section{Formalism}
\label{sec-formalism}

\subsection{Preliminaries}

We consider a massless self-interacting real scalar field $\phi$ with quartic coupling. 
The Lagrangian density is given by
\bea\label{lagrangian}
\mathcal{L} = \frac{1}{2} \partial^\mu \phi \partial_\mu \phi - \frac{g^2}{4!}\phi^4
\eea
where $g$ is the coupling constant. 
To mimic the kinematics of a high energy nuclear collision we work in Milne coordinates $(\tau,\eta,\vec{x}_{\perp})$ with 
\begin{eqnarray*}
    \tau &=& \sqrt{t^2-z^2} \\
    \eta &=& \frac{1}{2}\ln\left(\frac{t+z}{t-z}\right)\,.
\end{eqnarray*}
Under a Lorentz boost in the $z$-direction, the proper time is unchanged and $\eta$ is shifted by a constant. 
The  metric in Milne coordinates is
\bea
g_{\mu\nu} dx^\mu dx^\nu = d\tau^2-\tau^2 d\eta^2 - dx^2 -dy^2\,.
\eea
Figure 1 shows curves of constant $\tau$ and $\eta$.
\begin{figure}[H]
\centering
\includegraphics[scale=0.3]{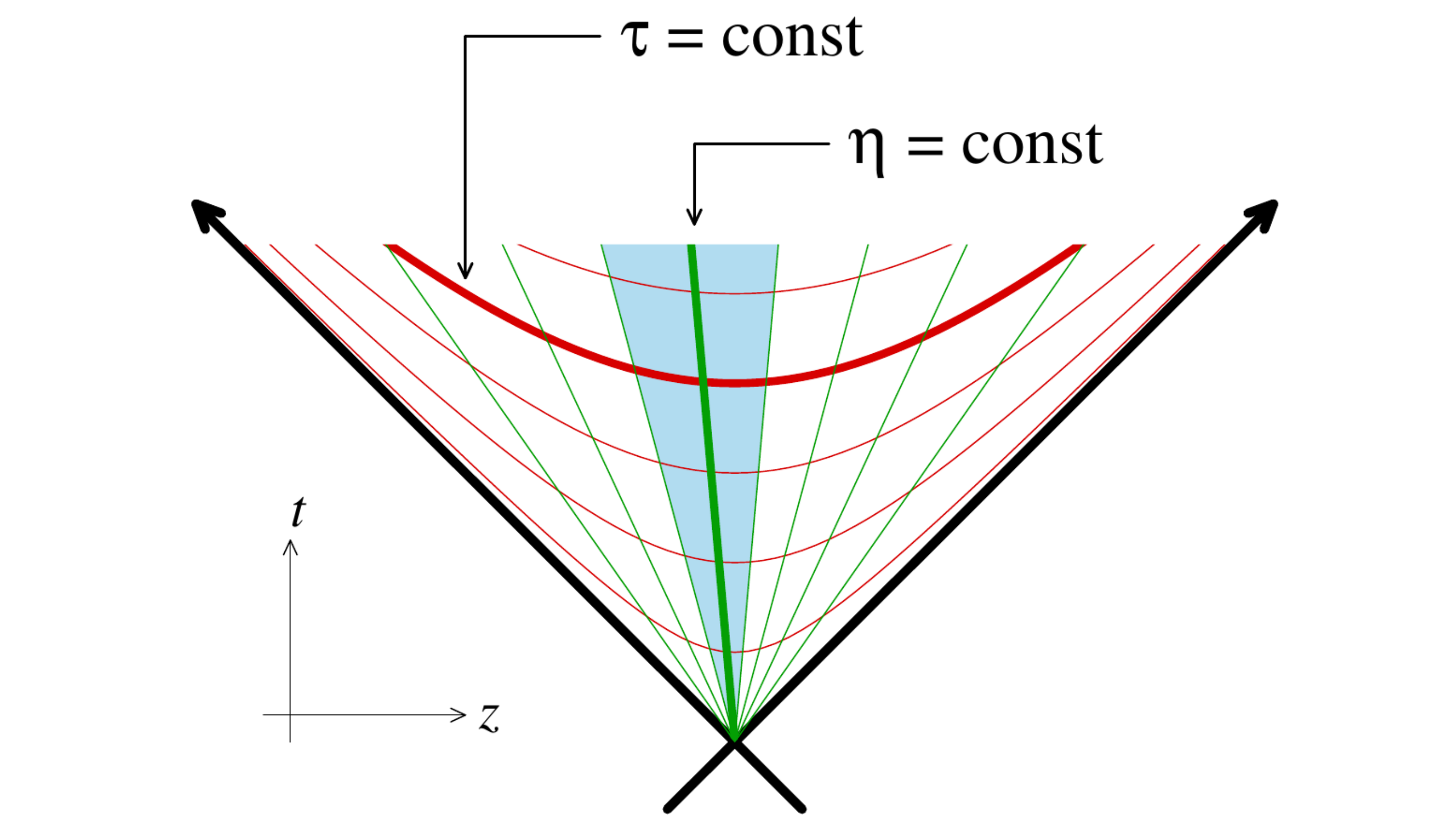} 
\caption{Representation of hypersurfaces of constant $\tau$ and $\eta$. The rapidity determines the location of  a particle along a surface of fixed $\tau$.}
\end{figure}

\subsection{The Resummation Procedure}

As explained in \cite{Dusling10,Dusling:2012ig}, observables calculated in a loop expansion exhibit  secular divergences at next-to-leading order that originate from instabilities of the classical solutions. Gelis {\it et. al.} propose to cure this problem using a resummation scheme that collects the leading secular terms at each order of an expansion in the coupling constant, by averaging over an ensemble of initial conditions. 
The energy-momentum tensor is ultraviolet divergent, but the divergence corresponds to a vacuum contribution and can be removed by repeating the calculation with the background field set to zero, and subtracting the results. This vacuum subtraction has been done for all the calculations presented in this paper. 

The equation of motion for the scalar field obtained from the Lagrangian (\ref{lagrangian}) is
\begin{equation}\label{eq-motion}
    \ddot{\phi}(\tau,\eta,\vec x_\perp)-\frac{1}{\tau}\dot{\phi} - \frac{1}{\tau^2}\partial^2_{\eta}\phi - \Delta_{\perp}\phi +\frac{g^2}{6}\phi^3 = 0
\end{equation}
where the ``dot'' indicates a derivative with respect to $\tau$, and $\Delta_{\perp}$ is the transverse Laplacian operator.
The initial field is written as the sum of a background field contribution,  $\varphi$, which is assumed boost invariant and therefore independent of $\eta$, and an $\eta$ dependent fluctuation which we call $\alpha$\footnote{The initial time $\tau_0$ is chosen to be small but nonzero  (see section \ref{sec-hankel} for further discussion).}
\bea
\phi^\chi(\tau_0,\eta,\vec x_\perp) = \varphi(\tau_0,\vec x_\perp) + \alpha^\chi(\tau_0,\eta,\vec x_\perp)\,.
\label{phi-0}
\eea
The initial background field $\varphi(\tau_0,\vec x_\perp)$ is discussed in section \ref{section-ic}. 
The index $\chi$ in equation (\ref{phi-0}) indicates that we have a
Gaussian ensemble of initial conditions defined as
\bea
&& \alpha^\chi(\tau_0,\eta,\vec x_\perp) = \int dK \big[c^\chi_K a_K+c^{\chi\,*}_K a_K^*\big]\,.\label{phi-1}
\eea
The index $K$ labels the momentum variables  $(\nu,\vec k_\perp)$ that are conjugate to the coordinate-space variables   $(\eta,\vec x_\perp)$, respectively. 
The notation $c^\chi_K$ indicates an element in a Gaussian distributed ensemble of $N_\chi$ random numbers, with variance
\bea
\langle c^*_K c_L\rangle = \frac{1}{2}\delta_{KL}\,.
\label{variance}
\eea 
We use the momentum space integration measure
\bea
 dK = \frac{d\nu}{2\pi} \frac{d \vec k_\perp}{(2\pi)^2}
\label{measure}
\eea
and the delta function in equation (\ref{variance}) is defined so that $\int dK \delta_{KL} = 1$.
The mode functions $a_K\equiv a_{\nu \vec k_\perp}(\tau_0,\eta,\vec x_\perp)$ are obtained from the linearized equations of motion
\bea
\ddot a_K + \frac{1}{\tau}\dot a_K -\frac{1}{\tau^2} \partial^2_{\eta} a_K -\Delta_\perp a_K +\frac{g^2}{2} \varphi^2(\tau_0,\vec x_\perp) a_K =0\,
\label{eqnx1}
\eea
and normalized so that 
$\int dK (a_K,a_L) = 1$ with 
\bea
(a_K,a_L) = i \tau \int d\eta \int d^2 \vec x_\perp \, \big(a_K^* \partial_\tau a_L - (\partial_\tau a^*_K)a_L \big)\,.
\label{inner}
\eea
Separating variables and performing the normalization one finds
\bea
a_K&\equiv& a_{\nu \vec k_\perp}(\tau_0,\eta,\vec x_\perp) = \frac{1}{2}\sqrt{\pi} e^{\pi\nu/2}\,e^{i\nu\eta}\chi_{\vec k_\perp}(\vec x_\perp)  H^{(2)}_{i\nu}(\lambda_{\vec k_\perp}\tau_0) \,\label{bkg}
\eea
where the  $\chi_{\vec{k}_\perp}$ is the solution of the eigenvalue equation
\bea
\left[-\Delta_\perp+\frac{g^2}{2} \varphi^2(\tau_0,\vec x_\perp)\right]\chi_{\vec k_\perp}(\vec x_\perp) = \lambda^2_{\vec k_\perp}\chi_{\vec k_\perp}(\vec x_\perp)\,.
\label{chi-1}
\eea

The field $\phi^\chi(\tau,\eta,\vec x_\perp)$ at finite proper time is obtained by solving equation (\ref{eq-motion}) with the initial condition $\phi^\chi(\tau_0,\eta,\vec x_\perp)$ obtained from equations (\ref{phi-0}, \ref{phi-1}, \ref{variance}, \ref{measure}, \ref{bkg}, \ref{chi-1}). 
From this point on we drop the subscript $\chi$.

\section{Observables}
\label{sec-observables}
\subsection{Energy momentum tensor}

The energy-momentum tensor of theory (\ref{lagrangian}) is 
\bea
T^{\mu\nu} = \partial^\mu \phi \partial^\nu\phi - g^{\mu\nu}\left[\frac{1}{2}\partial^\alpha\phi \partial_\alpha\phi - \frac{g^2}{4!}\phi^4\right]\,. \label{Tmunu}
\eea
The invariance of the Lagrangian under the conformal transformation 
\bea
g_{\mu\nu}\to \Omega^{-2} g_{\mu\nu};\quad\quad \phi\to \Omega^{-1}\phi
\eea 
implies that $T^{\mu\nu}$ is traceless on shell.

The expressions for the energy and pressure are
\bea
&& \epsilon = T^{00} = \frac{1}{2} \left((\partial_\tau\phi)^2+ \frac{(\partial_\eta \phi)^2}{\tau ^2}+(\partial_x \phi)^2+(\partial_y\phi)^2\right)+ V(\phi) \nn\\
&& p_L = \tau^2 T^{11} = \frac{1}{2} \left((\partial_\tau\phi)^2+ \frac{(\partial_\eta \phi)^2}{\tau ^2}-(\partial_x \phi)^2-(\partial_y\phi)^2\right)- V(\phi) \nn\\
&& p_T = \frac{1}{2}\left(T^{22}+T^{33}\right) =
\frac{1}{2} \left((\partial_\tau\phi)^2 -  \frac{(\partial_\eta \phi)^2}{\tau ^2} \right) - V(\phi) \,
\eea   
where $V(\phi)=g^2\phi^4/4!$. 
In terms of the energy and pressure, the trace condition is:
\bea
\epsilon= 2 p_T+ p_L\,.
\label{trace}
\eea

\subsection{Angular Momentum}
\label{sec-angularmomentum}

We use the standard Pauli-Lubanski formalism \cite{LUBANSKI1,LUBANSKI2} to obtain an expression for the angular momentum in terms of the energy-momentum tensor. 
We define the tensor field
\bea
M^{\mu\nu\lambda}=T^{\mu\nu}R^\lambda - T^{\mu\lambda}R^\nu \,
\label{M-def}
\eea
where $R^{\mu}$ is the coordinate vector. 
Using Stokes' theorem one obtains a set of six conserved quantities 
\bea
J^{\nu\lambda} = \int_\Sigma d^3 y \sqrt{|\gamma|}\; n_\mu M^{\mu\nu\lambda}\,,
\label{J1}
\eea
where $n^\mu$ is a unit vector perpendicular to the hypersurface $\Sigma$, $\gamma_{ij}$ is the induced metric on this hypersurface, and $d^3y$ is the corresponding volume element.
The angular momentum is obtained from the Pauli-Lubanski vector 
\bea
L_\mu = -\frac{1}{2}\epsilon_{\mu\alpha\beta\rho}J^{\alpha\beta}u^\rho
\label{L-def}
\eea
where $u^{\rho}$
is the vector that denotes the rest frame of the system. 
Equations (\ref{M-def}, \ref{J1}, \ref{L-def}) give 
\bea
L_\mu = -\frac{1}{2}\epsilon_{\mu\alpha\beta\rho}
\int d^3y \sqrt{\gamma}\,  n_\sigma \, u^\rho\,
(T^{\sigma\alpha}R^\beta - T^{\sigma\beta}R^\alpha)
\eea
where the energy-momentum tensor is given in equation (\ref{Tmunu}). 

To find the angular momentum on a surface of constant $\tau$ we define
\bea
n_\mu = \frac{\partial \tau}{\partial x^\mu}\,.
\eea
In Minkowski coordinates, this gives $n_\mu = (\text{cosh}(\eta),0,0,-\text{sinh}(\eta)$, and it easy to verify that 
$n^{\rm Milne}_\mu = (1,0,0,0)$, as expected. The fluid velocity is the local rest frame in comoving coordinates, which is written $u_{\rm Milne}^\rho = (1,0,0,0)$. In Minkowski coordinates this becomes $u^\rho = (\text{cosh}(\eta),0,0,\text{sinh}(\eta))$. 
We could calculate the angular momentum directly in Minkowski coordinates, or alternatively we could do the calculation in Milne coordinates and perform a coordinate transformation to obtain the Minkowski space result. We have checked our computations by verifying that both calculations give the same result. 
The components of the angular momenta about each of the Minkowski coordinate axes are
\bea
&& {L_t} = \tau\int d^2\vec{x}_\perp\, d\eta\, \text{sinh}(\eta)  \,\dot\phi \left(x \partial_y \phi -y \partial_x\phi \right) \nn  \\ 
&& {L_x} = \int d^2\vec{x}_\perp\, d\eta\,\dot\phi\, y \partial_\eta \phi \nn \\
&& {L_y} = -\int d^2\vec{x}_\perp\, d\eta\,\dot\phi\, x \partial_\eta \phi \nn \\
&& {L_z} = -\tau\, \int d^2\vec{x}_\perp\, d\eta\, \text{cosh}(\eta)\,\dot\phi\, \left(y \partial_x \phi -x \partial_y\phi \right) \,.\label{L-mink}
\eea
We note that all components of the angular momentum are dimensionless (in natural units, with $\hbar=1$).

\section{Numerical Implementation}
\label{sec-numerics}

\subsection{Lattice discretization}

We discretize  in both directions in the transverse plane with $L$ grid points and lattice spacing set to 1, which effectively means we define all dimensionful quantities in terms of the transverse lattice grid spacing.  
The rapidity variable $\eta$ is discretized with $N$ grid points and lattice spacing $h$. 
We consider a unit slice of rapidity, and therefore take $h=1/N$. 

The discretization of the transverse variables is straightforward. 
The discretized version of equation (\ref{chi-1}) is 
\bea
&& D_{ij;kl}\;\chi_{kl} = \lambda^2 \chi_{ij} 
\label{chi-2}
\eea
with
\bea
&& D_{ij;kl} = (4+V^{\prime\prime}_{ij})\delta_{ik}\delta_{jl} - (\delta_{i+1~k}+\delta_{i-1~k})\delta_{jl} - \delta_{ik}(\delta_{j+1~l}+\delta_{j-1~l})\,.
\label{D-mat}
\eea 
Since $D$ is a rank 4 tensor with $L^4$ components, we obtain $L^2$ eigenfunctions $\chi_{ij}^e$, and $L$ eigenvalues $\lambda^e$, with $e\in(1,L^2)$.
The normalized eigenfunctions are 
\bea
\sum_{ij} \chi^{*e}_{ij}\chi^{\bar e}_{ij} = L^2\,\delta^{e\bar e}
\label{eigen-norm}
\eea
and the momentum integration is discretized as
\bea
\int \frac{d^2\vec k_\perp}{(2\pi)^2} \to \frac{1}{L^2} \sum_{e=1}^{L^2}\,.
\label{dis-kperpint}
\eea 
Since the spatial lattice spacing is set to 1, an integral over transverse coordinates is discretized as 
\bea
\int d^2\vec x_\perp \to \sum_{i=1}^L\sum_{j=1}^L\,.
\eea

The discretization of the longitudinal variables is a little more subtle. 
%
%
The constraint 
\bea\partial_\eta^2 e^{i\nu\eta} = -\nu^2 e^{i\nu\eta} \eea
gives
\bea
\varepsilon^2_v := \nu^2 = \left(\frac{2}{h}\sin\left(\frac{\pi v}{N}\right)\right)^2\,
\eea
and we replace $\nu \to \varepsilon_v$ in every factor $e^{\pi\nu/2}$. For the complex exponential we use
$e^{i\nu\eta} \to e^{\frac{2\pi i v n}{N}}$.
The integral over $\nu$ becomes a sum over $v$ using
\bea
\int\frac{d\nu}{2\pi} \to \frac{1}{Nh} \sum_{v}^{N}\,.
\label{dis-nuint}
\eea

Combining these expressions we find the discretized versions of equations (\ref{phi-0}, \ref{variance}, \ref{bkg}): 
\bea
&& \alpha_{nij}(\tau) =  \frac{1}{N L^2 h} \sum_{v=1}^{N}  \sum_{p=1}^{L^2} \left[c_{vp} a^{vp}_{nij}(\tau)\, + \text{c.c.}\right]\, \nn \\
&& a^{vp}_{nij}(\tau) = \frac{1}{2}\sqrt{\pi} e^{\frac{2\pi i v n}{N}}\,\chi_{ij}^p \, e^{\pi\nu/2} H^{(2)}_{i\nu}(\lambda_{\vec k_\perp}\tau) \, \nn \\[2mm]
&& \langle c_{ve}c^*_{u\tilde e}\rangle = \frac{1}{2}N L^2 h \delta_{vu}\delta_{e\tilde e}\,.
\eea
To verify that discretization is done correctly we have checked  the discretized version of the normalization condition (\ref{inner}).

\subsection{Boundary conditions}

We use periodic boundary conditions, which means that the indices $(i,j)$ that correspond to the transverse spatial coordinates are defined modulo $L$, and the index $n$ for the rapidity is modulo $N$. 
The boundary conditions satisfy the self-adjointness condition
\bea
&& \nabla_F \phi(x) = \phi(i+1)-\phi(i) \nn \\
&& \nabla_B \phi(x) = \phi(i)-\phi(i-1) \nn \\
&& \sum_i f(i)\big(\nabla_F g(i)\big) = - \sum_i \big(\nabla_B f(i)\big) g(i) \,. \nn
\eea

\subsection{Hankel functions}
\label{sec-hankel}
The differential equation for the mode function was solved by separating variables, which gives the solution in (\ref{bkg}). 
The time dependent part of the equation is second order, and has two independent solutions which are the Hankel functions $H^{(1)}_{i\nu}(\lambda\tau)$ and $H^{(2)}_{i\nu}(\lambda\tau)$. We use only the second because it has positive frequency behaviour at large times
\bea
\lim_{\tau\to\infty}H^{(2)}_{i\nu}(\tau) = \sqrt{\frac{2}{\pi\tau}}e^{-i(\tau-i\pi\nu/2-\pi/4)}\,.
\eea
From now on we supress the superscript $(2)$ on the Hankel function. 
When $\tau\to 0$ the  Hankel function oscillates like $e^{\pm i \tau \nu}$ and the derivative diverges. 
Numerically we must start the evolution at a small positive time, which we choose as $\tau_0=10^{-2}$. 
One can check that the value chosen for this small initial time does not change the results at finite times. 

We describe below our method to calculate the Hankel functions. 
First we define the scaled function 
\bea
h_{i\nu}(\lambda\tau) = e^{\pi\nu/2}H_{i\nu}(\lambda\tau)
\eea
which is easier to calculate numerically. 
At large times one can obtain the scaled Hankel function for given values of $\nu$ and $\lambda$ from the asymptotic series
\bea
&& h_{i\nu}(\lambda\tau) = \sqrt{\frac{2}{\pi\lambda\tau}}e^{-i(\lambda\tau-\pi/4)} \sum_{k=0}^n t_k + {\cal O}(\tau^{-(n+1)})\nn\\
&& t_k = \frac{(-1)^k}{k!(2i\lambda\tau)^k}\prod_{s=1}^k\left(\nu^2+\frac{(2s-2)^2}{4}\right)\,.\label{hinkel-series}
\eea
This expression must be used carefully, because the series does not converge for arbitrarily large values $n$.
We proceed as follows. 
For a given value of $\nu$ and $\lambda$, choose some value of $\tau$ and look for a value of $k_{\rm max}$ so that $t_{k_{\rm max}+1}<10^{-9}$ and Max$(t_{k\le k_{\rm max}})<10^6$. If this $k_{\rm max}$ can be found, use equation (\ref{hinkel-series}) with $n=k_{\rm max}$. 
If $k_{\rm max}$ does not exist, then increase the chosen value of $\tau$ and try again. 
Using this procedure we can find $h_{i\nu}(\lambda\tau)$ and its first derivative for each value of $\nu$ and $\lambda$, for some (possibly very large) time. We then use adaptive fifth order Runge-Kutta  to find each Hankel function at the initial time $\tau_0$. 

\subsection{Discretized derivatives}

The conservation equation 
\bea
\frac{\partial \epsilon}{\partial\tau} + \frac{\epsilon+p_L}{\tau} = 0 \label{hydro}
\eea
is an exact equation that should be satisfied whether or not the system is in equilibrium. 
Also, we should have that the trace of the energy-momentum tensor is zero, so that equation (\ref{trace}) is satisfied.
%
It is easy to show analytically that these conditions are satisfied for background fields if we use forward derivatives: $\partial_x f(x) \to f(i+1)-f(i)$. 
The point is that while centered derivatives are not wrong,  much larger lattices must be used to achieve the same numerical accuracy. 

For angular momentum, the situation is different. 
All contributions to the angular momentum have an integral of the form $\int dx\,  \dot\varphi \, \partial_x\phi$. 
If the initial value of $\dot\varphi$ is constant, the integrand is a total derivative and therefore the integral will give zero. 
However, this is not well satisfied numerically with forward derivatives. 
In the calculation of angular momentum it is therefore better to use centered derivatives: $\partial_x f(x) \to (f(i+1)-f(i-1))/2$.

\subsection{Initial conditions}
\label{section-ic}

The initial conditions that we use for the background field and its derivative are
\bea
&& \varphi(\tau_0,i,j) = \varphi_0\,\cos(k_x i + k_y j )
\, \nn \\[2mm]
&& \dot\varphi(\tau_0,i,j) = \dot\varphi_0 \sin\left[\left(i-\frac{L+1}{2}\right)\frac{\pi}{L-1}\right]\,.
\label{init-c}
\eea
The argument of the sine function is $\mp\pi/2$ at $i=1$ and $i=L$, and zero at $i=(L+1)/2$, so the field has negative $\dot\varphi_0$ on the left side of the lattice and positive $\dot\varphi_0$ on the right side.

The astute reader will note that our initial classical field is not periodic, and therefore does not respect our boundary conditions. 
The reason is that we wish to avoid problems that may arise when resonant modes are considered, which in the present model would  correspond to the normal modes of the finite spatial lattice. 
For a large enough lattice, all modes are effectively periodic, and it is therefore expected that the precise form of the initialization is not important.

\section{Results and Discussion}
\label{sec-results}

All of our results are obtained with $L=41$ spatial grid points, $N=120$ points for the rapidity coordinate, and $N_\chi = 256$ configurations. The initial conditions for the background field are obtained from (\ref{init-c}) with $\varphi_0=15$, $k_x=k_y=1/\sqrt{2}$ and  $\dot\varphi_0=10$. 

\begin{figure}[H]
\centering
\includegraphics[scale=0.880]{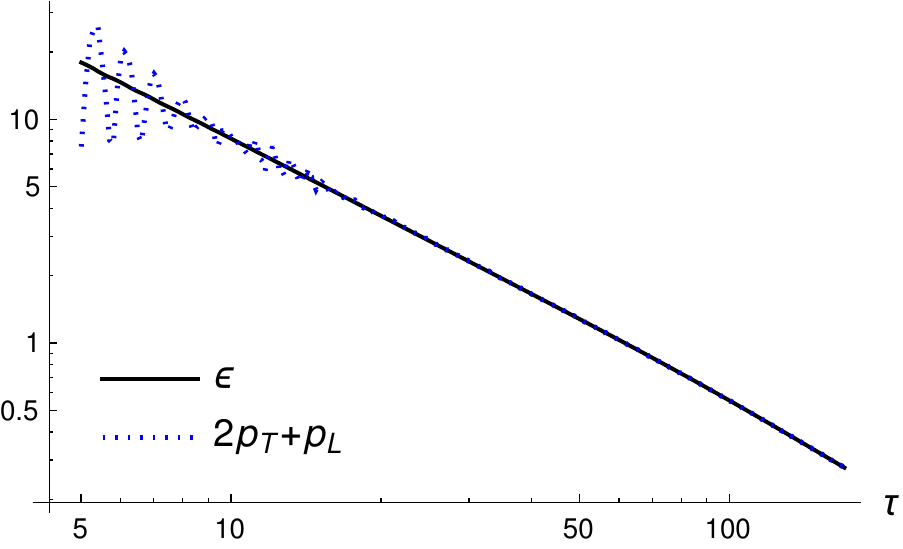} 
\caption{The energy and sum of the pressures as functions of $\tau$.\label{plot-eos}}
\end{figure}

\begin{figure}[H]
\centering
\includegraphics[scale=0.80]{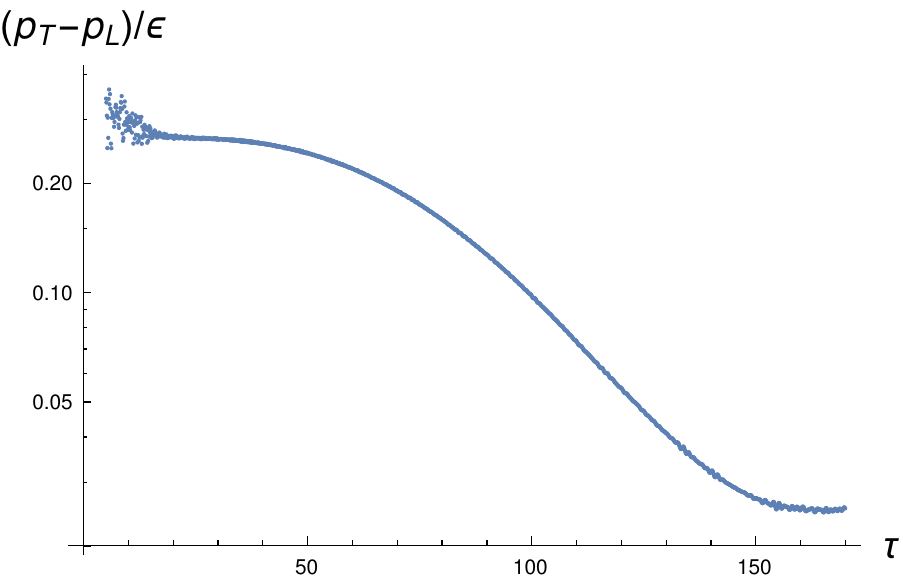} 
\includegraphics[scale=0.80]{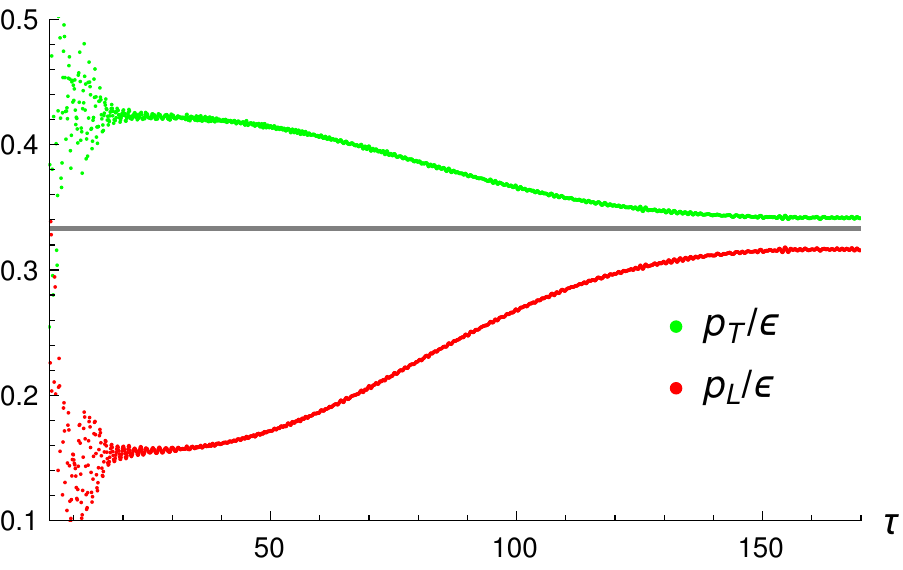} 
\caption{The transverse and longitudinal pressures, normalized by the energy density.\label{plot-iso}}
\end{figure}

To investigate if the system obeys equation (\ref{trace}) we compare the energy density and the sum of the pressures. This is shown in figure \ref{plot-eos}. One sees that after some initial oscillations have damped out, the condition $\epsilon = 2p_T+p_L$ is well satisfied. 
To see if the system approaches an isotropic state, and if it obeys an equation of state, we look at the transverse and longitudinal pressures. 
The left panel of figure \ref{plot-iso} shows that, after some initial oscillations have disappeared, the transverse and longitudinal pressures approach each other up to a time of about $\tau\approx 160$. 
The right panel shows the two pressures normalized by the energy density, both approaching 1/3, again up to $\tau\approx 160$.  
For large times, the simulation breaks down, which is not unexpected when one studies the dynamics of an expanding system inside a box of finite size. 

\begin{figure}[H]
\includegraphics[scale=1.80]{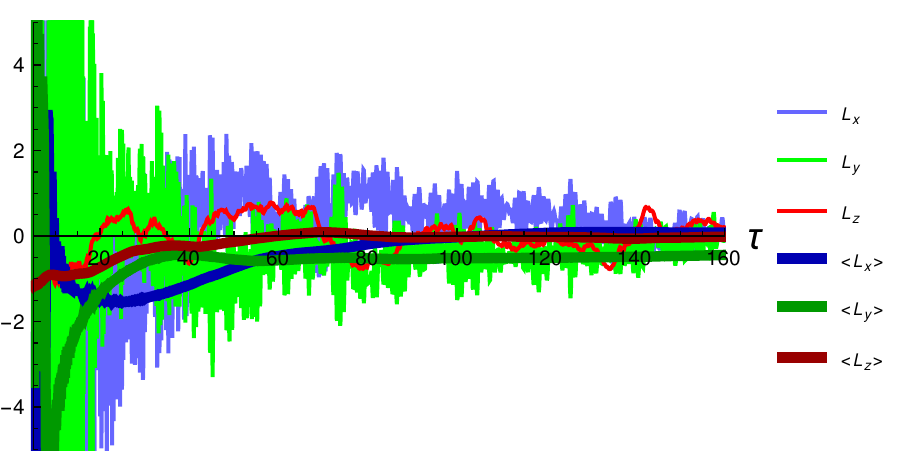} 
\caption{The three components of the angular momentum vector, and their accumulated averages. \label{plot-L}}
\end{figure}

In figure \ref{plot-L} we show the three components of the angular momentum in equation (\ref{L-mink}). 
The $z$-component, which depends weakly on the rapidity, is averaged over the unit slice of rapidity that we consider. 
In comparison with the energy and pressure, the oscillatory behaviour is more severe and does not completely disappear. To get a better idea of the overall behaviour, we also plot the accumulated average for each component, which is shown in figure \ref{plot-L} with the thick lines. In each case the darker colour corresponds to the average of the component with the same, but lighter, colour. The figure shows that even a fairly large initial angular momentum decays very quickly. 

\begin{figure}[H]
\centering
\includegraphics[scale=1.0]{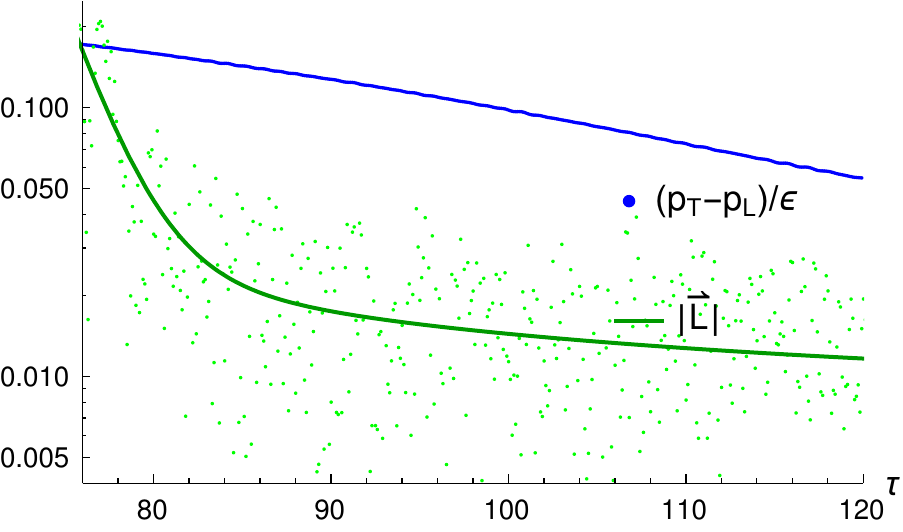} 
\caption{A comparison of $(p_T-p_L)/\epsilon$ (blue) and $|\vec L|$ (green), see text for details. \label{plot-L2}}
\end{figure}

We want to compare the time scales for the isotropization of the pressures, and the decay of the initial angular momentum. In figure \ref{plot-L2} we show in blue the curve in the left panel of figure \ref{plot-iso} over the range of $\tau$ for which the decay is strongest. 
To produce the light green points, we took the data for $|\vec L|$ versus $\tau$ with $\tau>12.0$, where the large initial fluctuations are mostly gone, and shifted the first point (which was (12.0, 9.10)) so that it sits on top of the first point of the data that made the blue curve. The dark green line is a fit obtained for this data using the function $A+B/\tau + Ce^{-D\tau}$. 
The plot shows clearly that the initial angular momentum decays much more quickly than the pressure anisotropy.

\section{Conclusions}
\label{sec-conclusions}
In this paper we have presented some preliminary results from our study of the angular momentum in an expanding system of rotating massless scalar fields. Our results indicate that even when a large amount of angular momentum is put into the system, it decays very rapidly. Future work will include an investigation of how much these results depend on the exact form of the initialization and the boundary conditions, and possibly the extension of the calculation to physical theories like QCD. 

\begin{acknowledgments}
Margaret Carrington gratefully acknowledges helpful discussions with Fran\c{c}ois Gelis. This work has been supported by the Natural Sciences and Engineering Research Council of Canada Discovery Grant program from grants 2017-00028 and 2018-04090. Marcelo Rubio acknowledges financial support provided under the European Union’s H2020 ERC Consolidator Grant ``GRavity from Astrophysical to Microscopic Scales'' grant agreement no. GRAMS-815673 at SISSA, Trieste. This research was enabled in part by support provided by WestGrid (www.westgrid.ca) and the Digital Research Alliance of Canada (alliancecan.ca).

\end{acknowledgments}

\bibliography{refs}

\end{document}